\documentclass[sigconf,natbib]{acmart}

\usepackage{multirow}
\usepackage{tabularx,booktabs}
\usepackage{amsmath}
\usepackage{amsfonts}
\usepackage{amsbsy}
\usepackage{enumitem}
\usepackage{comment}
\usepackage{makecell}
\usepackage{soul}

\usepackage{xcolor}
\usepackage{colortbl}
\definecolor{graylite}{gray}{.89}
\newcolumntype{g}{>{\columncolor{graylite}}c}

\AtBeginDocument{%
  \providecommand\BibTeX{{%
    \normalfont B\kern-0.5em{\scshape i\kern-0.25em b}\kern-0.8em\TeX}}}

\begin{document}

\title{Zero-Shot Next-Item Recommendation using Large Pretrained Language Models}

\author{Lei Wang}
\email{lei.wang.2019@phdcs.smu.edu.sg}
\author{Ee-Peng Lim*}\thanks{*Corresponding Author.}
\email{eplim@smu.edu.sg}
\affiliation{%
  \institution{Singapore Management University}
  \country{Singapore}
}

\renewcommand{\shortauthors}{Lei Wang, Ee-Peng Lim}

\begin{abstract}

Large language models (LLMs) have achieved impressive zero-shot performance in various natural language processing (NLP) tasks, demonstrating their capabilities for inference without training examples.
Despite their success, no research has yet explored the potential of LLMs to perform next-item recommendations in the zero-shot setting.
We have identified two major challenges that must be addressed to enable LLMs to act effectively as recommenders.
First, the recommendation space can be extremely large for LLMs, and LLMs do not know about the target user's past interacted items and preferences.
To address this gap, we propose a prompting strategy called \textbf{Zero-Shot Next-Item Recommendation (NIR)} prompting that directs LLMs to make next-item recommendations.
Specifically, the NIR-based strategy involves using an external module to generate candidate items based on user-filtering or item-filtering.
Our strategy incorporates a 3-step prompting that guides GPT-3 to carry subtasks that capture the user's preferences, select representative previously watched movies, and recommend a ranked list of 10 movies.
We evaluate the proposed approach using GPT-3 on MovieLens 100K dataset and show that it achieves strong zero-shot performance, even outperforming some strong sequential recommendation models trained on the entire training dataset. These promising results highlight the ample research opportunities to use LLMs as recommenders. The code can be found at \href{https://github.com/AGI-Edgerunners/LLM-Next-Item-Rec}{https://github.com/AGI-Edgerunners/LLM-Next-Item-Rec}.
\end{abstract}




\maketitle

\section{Introduction}
\label{sec:intro}

Large language models (LLMs)~\cite{gpt3, opt, chowdhery2022palm}, such as GPT-3~\cite{gpt3}, have achieved impressive results in various natural language processing (NLP) tasks. Nevertheless, LLMs are also very large and often accessible only via some API service.  
Hence, they cannot be fine-tuned like the earlier pre-trained language models (PTMs)~\cite{devlin2018bert, radford2019language}.
Many works have demonstrated that LLMs are capable of solving many known NLP problems through task-specific prompts under the zero-shot setting, i.e., without any demonstration examples or further training~\cite{gpt3, chowdhery2022palm}. 
Nevertheless, using LLMs to perform next-item recommendations in the zero-shot setting is still a research topic in the nascent stage.

\begin{figure}[t]
    \centering
    \includegraphics[width=1.0\linewidth]{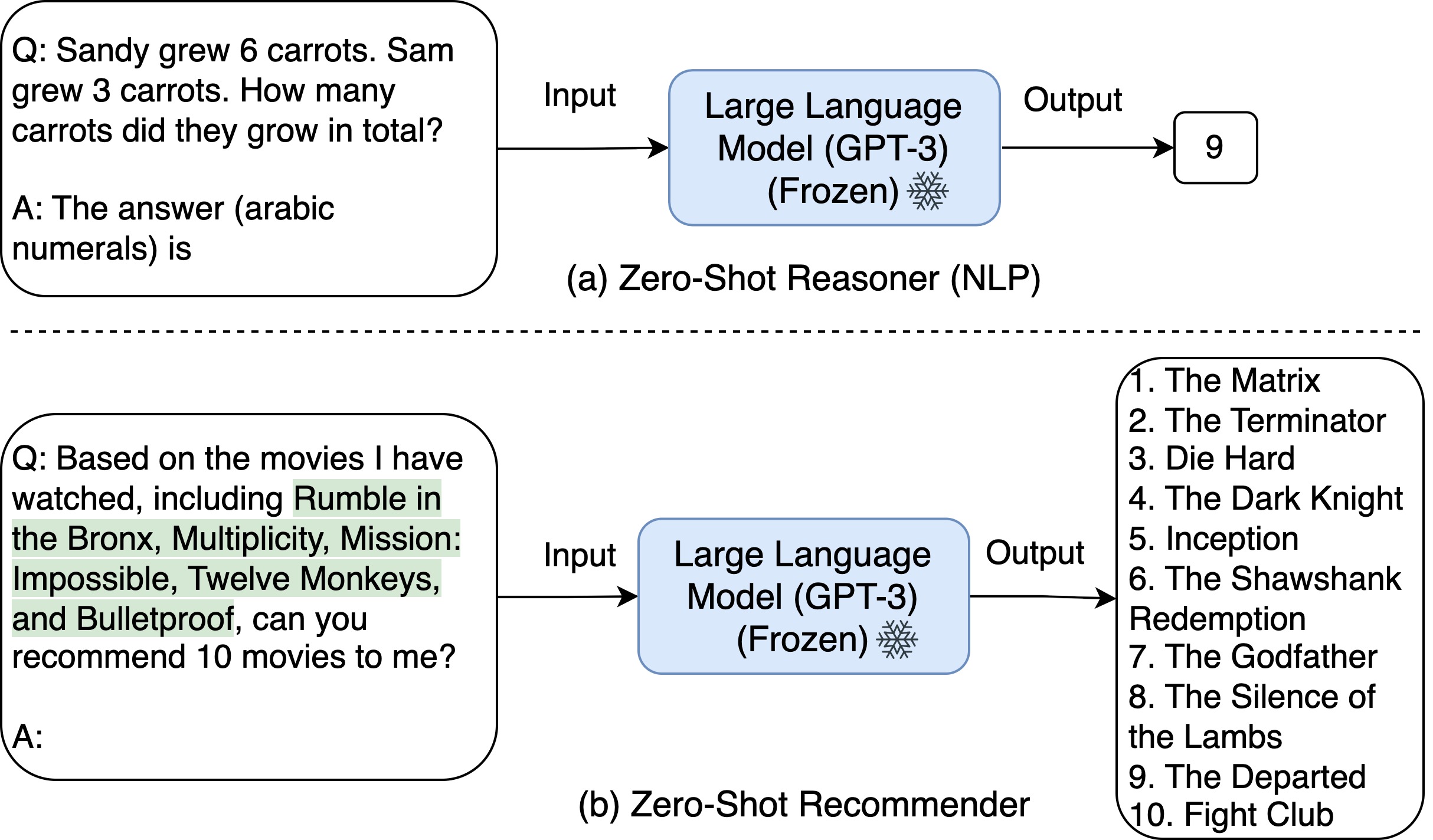}
    \caption{Example inputs and outputs of GPT-3 with zero-shot prompting for a NLP task and a recommendation task.}
    \label{fig:example}
\end{figure}

We use Figure~\ref{fig:example} to illustrate the differences between a NLP reasoning task and a recommendation task. This NLP reasoning task provides GPT-3~\cite{kojima2022large} (also the default LLM in this work) 
a question in a prompt and the latter generates the answer text (e.g., ``9'').
Unlike this NLP task which can directly rely on built-in textual knowledge of LLMs, the recommendation task requires LLMs to know the target user's previous item-interactions, the universe of items to be recommended, and the appropriate approach to select the recommended items.  Given that LLMs are not naturally trained to perform recommendation, poor results are expected when directly using them to perform recommendation~\cite{zhang2021language}. Moreover, LLMs can only contribute to recommendation when they have some background knowledge about the items to be recommended.  For recommendation of items in proprietary domains, it is unclear how LLMs can be of much use.   

In our research, we therefore assume that items for recommendation should appear in training data of LLMs.  Examples of such items include movies, songs, novels, online games, etc..  For illustration and evaluation purposes, we focus on the next-movie recommendation task.  We also choose GPT-3 as the LLM due to its popularity and accessibility. 
As depicted in Figure~\ref{fig:example}(b), we show a simple prompting strategy that directly incorporates user's previously watched movies into a text prompt, i.e., ``Based on the movies I have watched, including...'', followed by a question ``can you recommend 10 movies to me?''.  While this prompting strategy allows a GPT-3 to act as a movie recommender, its recommendation accuracy is likely to be poor due to an \textit{extremely large recommendation space} and \textit{inadequate user preference modeling}.

In this paper, we therefore propose a principled approach to the next-item recommendation called \textbf{Zero-Shot Next-Item Recommendation (NIR) prompting}, which involves a 3-step prompting strategy that significantly outperforms simple prompting in the zero-shot setting. Zero-Shot NIR adopts a three-pronged approach to enhance recommendation accuracy.  First, it restricts the recommendation space to the scope of MovieLens~\cite{harper2015movielens} dataset 
by deriving a candidate movie set for the target user using the user filtering and item filtering techniques well known in previous next-item recommendation research. 
Second, Zero-Shot NIR prompting strategy performs multi-step prompting of GPT-3 to capture the user's preferences (Step 1), select representative movies from the user's previously watched movies (Step 2), and recommend a ranked list of 10 movies (Step 3). 
Finally, Zero-Shot NIR introduces a formatting technique in Step 3 to facilitate the extraction of movie items from the answer text generated by GPT-3 (i.e., ``a watched movie: <- a candidate movie ->''). 
We evaluate our approach using MovieLens 100K and GPT-3 engine \texttt{text-davinci-003}. 
The experimental results show that Zero-Shot NIR prompting achieves good recommendation accuracy in the zero-shot setting and is comparable to other next-item recommendation methods trained using a large dataset. 

\section{Related Works}

Next-item recommendation is an important and well studied research problem.  Early research works proposed Markov Chains to model low-order relationships between items for next-item recommendation ~\cite{rendle2010factorizing, he2016fusing}. 
With the advancement of neural models, deep neural networks~\cite{hidasi2015session, tang2018personalized, kang2018self, huang2018improving, wang2020next, sun2019bert4rec, chang2021sequential} have been applied to the modeling of sequential patterns which leads to improved recommendation accuracy. 
Recent research has also explored the use of data augmentation and contrastive learning to enhance the representations of users and items, thereby making further improvement to recommendation performance~\cite{zhou2020s3, xie2020contrastive,  liu2021self, yao2021self, wu2021self}. 
Nevertheless, all the above methods require model training using users' historical item-interactions. In other words, they are not capable of making recommendations in the zero-shot setting.  To the best of our knowledge, there has been very little research on zero-shot recommendation, and it remains to be unclear whether LLMs can be good recommenders.

Among the earlier efforts in LLM-based recommendation~\cite{li2022personalized, zhang2021language, sileo2022zero, cui2022m6, geng2022recommendation, wu2022personalized}, ~\citet{zhang2021language} proposed to use GPT-2~\cite{radford2019language} or BERT~\cite{devlin2018bert} as the backbone recommender, making the next movie prediction based on five previously watched movies by the target user. 
However, the huge recommendation space and inadequate user preference modeling make it perform poorly.
With newer LLMs such as GPT-3~\cite{gpt3}, OPT~\cite{opt}, and PaLM~\cite{chowdhery2022palm} which have shown significantly improved results in various NLP tasks, our work chooses GPT-3 to be the LLM for developing more effective zero-shot recommendation methods.
Instead of designing the prompting strategy from scratch, our proposed Zero-Shot NIR prompting strategy incorporates user and item filtering approach to derive a candidate movie set and devises a 3-step prompting approach. This way, it mimics well-known recommendation techniques to achieve more accurate zero-shot recommendations. 

\section{Zero-Shot NIR Prompting Strategy}
\label{sec:nir}

\subsection{Overview}

\begin{figure}[t]
    \centering
    \includegraphics[width=1.0\linewidth]{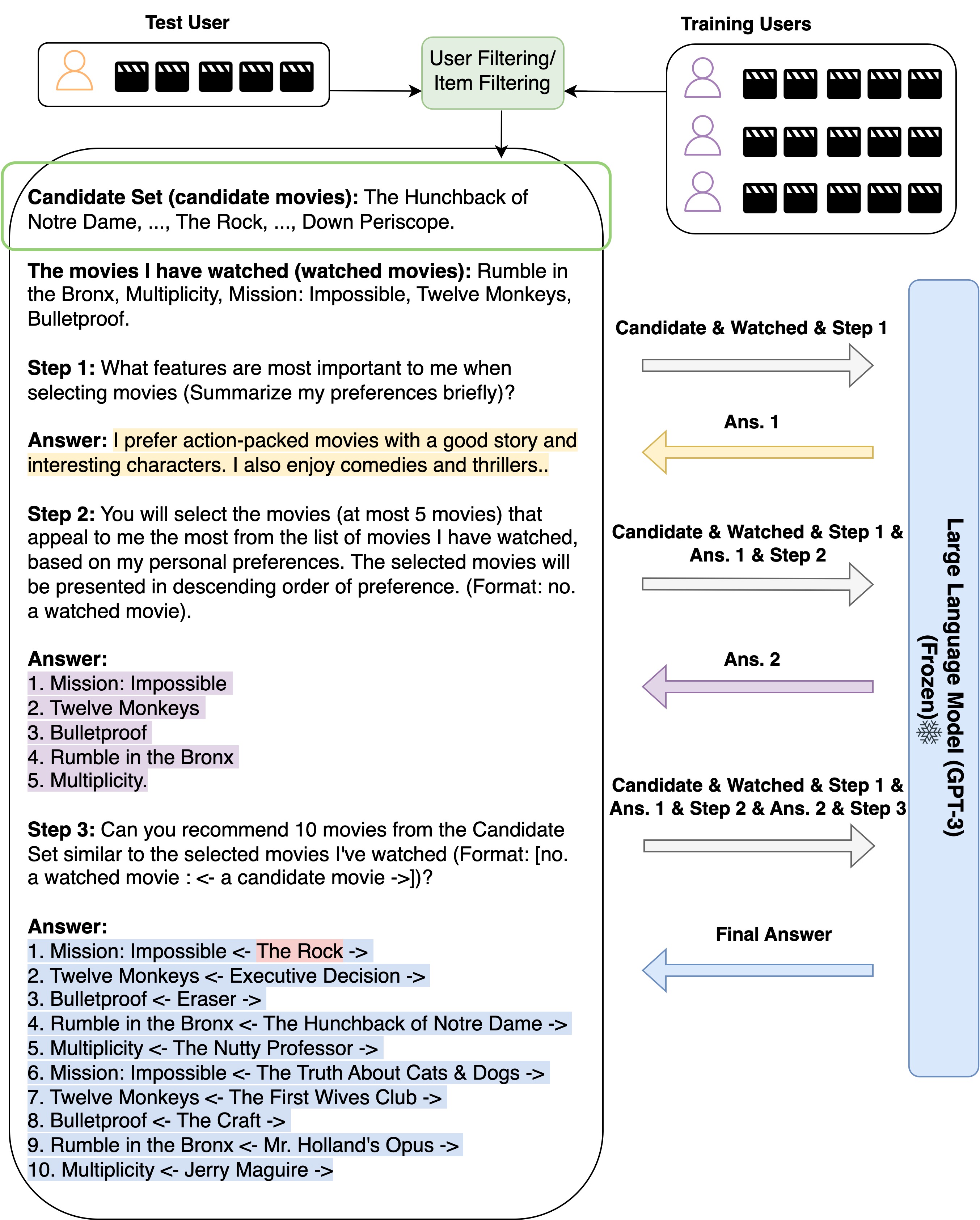}
    \vspace{-5pt}
    \caption{Zero-Shot NIR prompts. The ground truth movie (i.e., The Rock) has been highlighted in {\color{red} red}.}
    \label{fig:approach}
    \vspace{-5pt}
\end{figure}

Zero-Shot NIR prompting is a multi-step prompting strategy that enables GPT-3 to act as a next-item recommender in the zero-shot setting.  
Figure~\ref{fig:approach} illustrates the entire process of our Zero-Shot NIR prompting strategy.  It consists of three components: 
\begin{itemize}[leftmargin=*]
    \item \textbf{Candidate set construction:}  This component uses user filtering or item filtering to create a candidate set for each target user, so as to narrow down the recommendation space. These candidate movies are then used to build the three-step GPT-3 prompts.
    \item \textbf{Three-step GPT-3 prompting:}  This component involves three instruction prompts corresponding to three subtasks. In the first subtask (\textit{user preference subtask}), we design a user preference prompt to probe GPT-3 to summarize the user's preferences based on the previously interacted movies by the target user. In the second subtask (\textit{representative movies subtask}), we create a prompt that combines the user preference prompt with the prompt answer as well as a trigger instruction to request GPT-3 to select representative movies in descending order of preference. 
    In the third subtask, we integrate the representative movies prompt, its answers, and a question to create the recommendation prompt to guide GPT-3 to recommend 10 movies from the candidate movie set that are similar to the representative movies. The result is expected to be in the following format: ``a watched movie: <- a candidate movie ->''.

    \item \textbf{Answer extraction:}
    This component extracts the recommended items from the textual results of three-step GPT-3 prompting using a simple rule-based extraction method.  The extracted recommended movie results can be used in downstream applications or for performance evaluation.
\end{itemize}

\subsection{Candidate Set Construction}

As mentioned in Section~\ref{sec:intro}, an extremely large recommendation space poses a major challenge to LLM-based recommendation.  An unconstrained recommendation space can complicate both practical use and evaluation of recommendation results. On the other hand, it is infeasible to feed the LLM with all items expecting the former to be aware of the item universe.  Even for the purpose performance evaluation, the MovieLens 100K dataset contains a set of 1,683 movies which is still too large to fit into a prompt. 

In our Zero-Shot NIR prompting, we therefore propose to first construct the candidate movie set for each target user using a principled approach. The candidate movies should satisfy two criteria: (a) they should be more relevant to the user instead of randomly selected movies, and (b) the size of candidate movies should be small so as to fit into a prompt. 
To achieve these criteria, we use two well known principles for determining candidate movies, i.e., \textit{user filtering} and \textit{item filtering}.

\textbf{User Filtering. } This principle assumes that the candidate movies should also be liked by other users similar to the target user.  Hence, we first represent every user by a multi-hot vector of their watched movies.  Users similar to the target user are then derived by cosine similarity between the target user's vector and vectors of other users. 
Next, we select the $m$ most similar users and the candidate movie set of size $s$ is constructed by selecting the most popular movies among the interacted movies of similar users.

\textbf{Item Filtering.} 
Similar to user filtering, we represent each movie by a multi-hot vector based on its interacted users. 
Using cosine similarity between two movies, we select the $n$ most similar movies for each movie in the target user's interaction history.
We then generate a candidate set of size $s$ based on the ``popularity'' of these similar movies among the movies in the target user's interaction history.

The candidate set construction can be pre-computed by an external module. We then incorporate the candidate movies into the subsequent prompts for recommendation using the sentence: ``Candidate Set (candidate movies):'' as shown in Figure~\ref{fig:approach}.  Following the candidate set, the prompts also include the list of target user's previously interacted movies starting with: ``The movies I have watched (watched movies):''.

\subsection{Three-Step Prompting}

Figure~\ref{fig:example} (b) shows that the simple prompting method overlooks candidate movies, user preferences, and previously interacted movies that best reflect the user's tastes. 
In contrast, our proposed three-step prompting approach uses three rounds of prompting to perform three subtasks: capturing the target user's preferences, ranking previously interacted movies by user's preferences, and recommending 10 similar movies from the candidate set. 


\noindent \textbf{Step 1: User Preference Prompting.} To capture the user's preferences, we include the sentence  ``Step 1: What features are most important to me when selecting movies (summarize my preferences briefly)?'' into the first prompt. 
As shown in Figure~\ref{fig:approach}, the answer returned by GPT-3 summarizes the target user preference (highlighted in {\color{yellow} yellow}).

\noindent \textbf{Step 2: Representative Movie Selection Prompting.} As the second step, this prompt includes the previous prompt text appended with the answer of Step 1. It then includes the instruction: ``Step 2: You will select the movies ... that appeal to me the most ... presented in descending order of preference (...)'' to determine the previously interacted movies that best reflect the target user's tastes.
Figure~\ref{fig:approach} shows the GPT-3's answers highlighted in {\color{purple} purple}.

\noindent \textbf{Step 3: Recommendation Prompting.} Again, this prompt includes the previous text appended with the answers of 
 Step 2. It then includes the instruction ``Step 3: Can you recommend 10 movies from the Candidate Set similar to ...''. This prompt explicitly instructs GPT-3 to generate 10 recommended movies from the candidate set as highlighted in {\color{blue} blue}.

\subsection{Answer Extraction}

We add the hint ``(Format: ... <- a candidate movie ->])'' to the third prompt to generate answers in the desired format for easy extraction.  Our study has shown that such a hint has worked very well. 

\section{Experiments}

\subsection{Experimental Setup}

\noindent \textbf{Dataset.} The proposed prompting approach is evaluated on a widely-used movie recommendation dataset MovieLens 100K~\cite{harper2015movielens}, which contains 944 users and 1,683 movies.

\noindent \textbf{Baselines.} We compare our proposed NIR prompting strategy with two types of baselines: \textit{strong next-item recommendation baselines} and \textit{zero-shot baselines}. 
The former includes \textbf{POP} (a popularity-based model), \textbf{FPMC}~\cite{rendle2010factorizing} (an approach combining matrix factorization and Markov chains), \textbf{GRU4}~\cite{hidasi2015session} (a GRU-based sequential recommendation model), \textbf{SASRec}~\cite{kang2018self} (a sequential recommendation model with self-attention), and \textbf{CL4SRec}~\cite{xie2020contrastive} (a contrastive learning based sequential recommendation model).  As these strong baselines have the advantage of full model  training, they are expected to outperform zero-shot methods. 
The zero-shot baselines include \textbf{Zero-Shot Simple Prompting}, \textbf{CS-Random-IF} (that randomly selects 10 movies from the item filtering-based candidate set), and \textbf{CS-Random-UF} (that randomly selects 10 movies from the UF-based candidate set). 
We implement our NIR prompting strategy with four variants. 
\textbf{NIR-Combine-IF/NIR-Combine-UF} combines the 3 steps into a single prompt leaving out the intermediate answers. We only prompt GPT-3 once to generate 10 recommended movies from the IF/UF-based candidate set. 
\textbf{NIR-Multi-IF/NIR-Multi-UF} uses three separate prompts to guide GPT-3 step-by-step and to incorporate intermediate answers to the subsequent prompts (as shown in Figure~\ref{fig:approach}) with the IF/UF-based candidate set.

\noindent \textbf{Implementations.} 
In our experiments, we employ the public GPT-3 \texttt{text-davinci-003} (175B) as the backbone language model, one of the most popular LLMs with public APIs\footnote{https://beta.openai.com/docs/models/gpt-3}. To ensure consistent output, we set the temperature parameter to 0. For evaluation, we adopt the same evaluation metrics as in CL4SRec and report HR@10 and NDCG@10 for all methods.

\subsection{Main Results}

\begin{table}[t]\centering


\setlength{\tabcolsep}{7pt}
\caption{Main result comparison on MovieLen 100K.
}
\small
\begin{tabular}{llcc}\toprule
\multirow{ 2}{*}{Setting}& \multirow{ 2}{*}{Method} &\multicolumn{2}{c}{MovieLens 100K}\\
 & &HR@10   &NDCG@10  \\\midrule

\multirow{ 4}{*}{Full Training}& POP & {0.0519}  & {0.0216}\\

& FPMC   &  {0.1018}  & {0.0463}\\

& GRU4Rec &  {0.1230}  & {0.0559}\\

& SASRec  & 0.1241  & 0.0573 \\


& CL4SRec  & 0.1273  & 0.0617 \\

\midrule

\multirow{ 7}{*}{Zero-Shot} & Simple Prompting & {0.0297}   & {0.0097} \\
& CS-Combine-IF & {0.0805}  & {0.0352}\\
&  CS-Combine-UF & {0.0954}   & {0.0457} \\
& NIR-Single-IF & {0.0975}   & {0.0501} \\
& NIR-Single-UF & {0.1135}   & {0.0529} \\
& NIR-Multi-IF & {0.1028}   & {0.0505} \\
& NIR-Multi-UF & {0.1187}   & {0.0546} \\

\bottomrule
\end{tabular}
\label{tab:main_results}
\end{table}

As shown in Table~\ref{tab:main_results}, 
our NIR-based methods (NIR-Single-IF, NIR-Single-UF, NIR-Multi-IF, NIR-Multi-UF) outperform POP baseline by a large margin. Interestingly, NIR-Single-UF, NIR-Multi-IF, and NIR-Multi-UF consistently outperform FPMC, a fully trained method. 
Compared with very strong sequential recommendation models (i.e., GRU4Rec, SASRec, and CL4SRec), the three NIR-based methods still deliver slightly worse but competitive performance, suggesting that LLMs with proper prompting strategy can be reasonably good zero-shot recommenders. 

In the zero-shot setting, CS-Random-UF(IF)'s superior performance over Simple Prompting show that candidate set not only reduce the recommendation space, but also improve performance. 
Our proposed NIR-based prompting methods consistently outperforms Simple Prompting and CS-Random-IF/UF, suggesting that incorporating user preference, representative movie selection, and formatting techniques in the prompting process allow GPT-3 to make better recommendations.
As Multi-IF(UF) ouperforms Combine-IF(UF), we know that separate prompts incorporating the intermediate answers into subsequent prompts leads to more accurate recommendations. 
Finally, UF-based HIR-based prompting consistently outperforms IF-based prompting, indicating that UF yields better candidate sets than IF.

\subsection{Detailed Analysis}

\textbf{Effects of Components of Prompting.}
We now conduct an ablation study on NIP-Multi-UF to evaluate the contribution of different components of our prompting strategy.
Table 2 shows that all prompting components contribute to recommendation performance. 
Simple Prompting method with a candidate set (HR@10=0.1071) outperforms that without a candidate set (HR@10=0.0297).  
Incorporating user preferences or representative movie selection into the prompting improves performance, indicating that task-specific instructions can guide GPT-3 to perform better.

\begin{table}
\centering
\caption{
 Ablation study of the impact of different components in the proposed prompting on MovieLens 100K.
}
\scalebox{0.84}{
\begin{tabular}{ccc|c}
\toprule

\multirow{1}{*}{Candidate Set}&
\multirow{1}{*}{User Preference}& 
\multirow{1}{*}{Representative Movies}&
 HR@10\\

\midrule
  \textendash & \textendash &  \textendash & {0.0297}   
 \\
   \checkmark &\textendash & \textendash & 
 {0.1071}   \\
 \checkmark &\checkmark & \textendash & 
{0.1136} 
  \\
 \checkmark &\textendash & \checkmark & {0.1082}   \\
 \checkmark &\checkmark & \checkmark & {0.1187}  \\

\bottomrule
\end{tabular}
}

\label{tab:ablation_study}
\end{table}

\begin{figure}[t]
    \centering
    \includegraphics[width=0.6\linewidth]{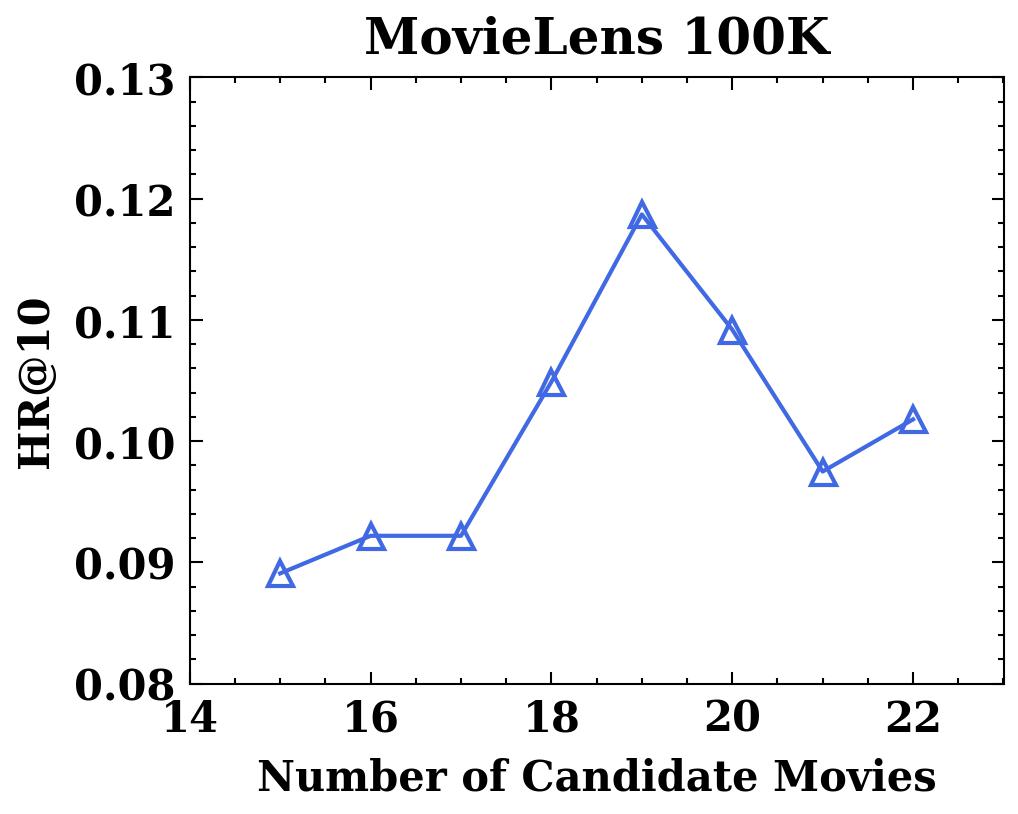}
    \caption{Results for different candidate set sizes.}
    \label{fig:cand_size}
\end{figure}

\textbf{Impact of Candidate Set Size.}
We investigate the impact of candidate movie number by varying the set size from 15 to 22 while keeping other parameters unchanged. The results in Figure~\ref{fig:cand_size} indicate that recommendation performance is sensitive to candidate set size. The best results occur when there are 19 candidate movies, and smaller or larger set size causes performance degradation. One possible explanation is that a small candidate set restricts the performance limit, while a large set increases the difficulty of making accurate recommendations using GPT-3.

\section{Conclusion}

In this paper, we propose a three-step prompting strategy called Zero-Shot Next-Item Recommendation (NIR) for GPT-3 to make next-movie recommendations without further training. We evaluate our approach on a movie recommendation dataset and demonstrate its strong zero-shot performance. Our results highlight the potential of using LLMs in zero-shot recommendation and call for further exploration of using LLMs in recommendation tasks.  This work can extended in several directions, including recommendation in other domains and few-shot setting. 

\bibliographystyle{ACM-Reference-Format}
\bibliography{sample-base}

\end{document}